\documentclass[aps,prapplied,preprint,groupedaddress,floatfix]{revtex4-1}
\usepackage{hyperref,graphicx,bm}

\begin{document}

\title{Spin orientation and magnetostriction of Tb\textsubscript{1-x}Dy\textsubscript{x}Fe\textsubscript{2}
from first principles}
\author{Christopher E.\ Patrick}
\affiliation{
Department of Materials, University of Oxford,
Parks Road,
Oxford OX1 3PH, UK}
\email{christopher.patrick@materials.ox.ac.uk}
\author{George A.\ Marchant}
\author{Julie B. Staunton}
\affiliation{Department of Physics, University of Warwick,
Coventry CV4 7AL, UK}

\date{\today}

\begin{abstract}
The optimal amount of dysprosium in the highly magnetostrictive
rare-earth compounds Tb$_{1-x}$Dy$_x$Fe$_2$ 
for room temperature applications has long been known to be
$x$=0.73 (Terfenol-D).
Here, we derive this value from first principles by calculating 
the easy magnetization direction and magnetostriction as a 
function of composition and temperature.
We use crystal field coefficients obtained within density-functional
theory  to
construct phenomenological anisotropy
and magnetoelastic constants.
The temperature dependence of these constants is obtained
from disordered local moment calculations of the rare earth magnetic order
parameter.
Our calculations 
find the critical Dy concentration required
to switch the magnetization direction at room temperature
to be $x_c$=0.78, with magnetostrictions
$\lambda_{111}$=2700 and $\lambda_{100}$=-430~ppm,
close to the Terfenol-D values.
\end{abstract}

\maketitle 

\section{Introduction}

The cubic Laves phase compound Terfenol-D (Tb$_{1-x}$Dy$_x$Fe$_2$, $x$ = 0.73) has
unparalleled magnetostrictive properties at room temperature,
developing strains of 1600~ppm when a small
magnetic field is applied and rotated between the [100] and [111] crystal
directions~\cite{Abbundi1977,Clark1973,Clark1977}.
Originally developed for sonar~\cite{Tremoletbook},
Terfenol-D has a range of potential applications, including
vibrational energy harvesting~\cite{Staley2005,Deng2017},
non-destructive testing~\cite{Rudd2018}
and multiferroic devices~\cite{Wang2017}.
The latter concept couples magnetostrictive and piezoelectric 
materials to control electric polarization (or magnetization)
with a magnetic (or electric) field, essential for
magnetic sensors or magnetoresistive memory~\cite{Li20182}.

While remarkable for its magnetostriction, Terfenol-D
does suffer from two drawbacks: it is brittle~\cite{Atulasimha2011}
and, due to its reliance on the critical heavy rare earths (REs) Tb and Dy,
it is expensive~\cite{Vincent2020}.
Intense research has been aimed at finding new materials
with reduced or zero RE content and better mechanical
properties, with the notable successes of Fe-Ga and Fe-Al 
(Galfenol and Alfenol)~\cite{Clark2003,Clark2008}.
Computational modelling, adopting a
first-principles (parameter-free) methodology, provides a complementary
approach to experimentally searching for new materials, as
well as understanding 
existing ones~\cite{Brooks19912,Richter1998,Buck1999,Gavrilenko2001,Fahnle2002,Wang2013}.
However, despite Terfenol-D's huge importance as a magnetostrictive material,
first-principles modelling has not yet been able to answer 
a basic question, namely: why is the optimum dysprosium content $x$=0.73?

Experimentally, the question can be answered by considering
the spin orientation phase diagram~\cite{Atzmony1973}, which
maps out the preferred (easy) direction of magnetization 
of Tb$_{1-x}$Dy$_x$Fe$_2$ as a function of $x$ and temperature $T$.
At $T$=300~K, for $x$ $\leq$ 0.6 the easy direction is along [111];
for $x$ $\geq$ 0.9, it is [100].
The critical concentration $x_c$=0.73 lies within the
soft boundary between these two regions of the phase diagram, and
corresponds to a low magnetocrystalline anisotropy (MCA).
The low MCA is essential for the aforementioned applications, 
since then only a small
field is needed to trigger a magnetostrictive response.
It is also important to note that this critical concentration $x_c$
reduces with temperature~\cite{Atzmony1973}.

A first-principles understanding of Terfenol-D 
therefore requires calculating the spin orientation diagram.
These have been calculated in the past using crystal field
(CF) theory~\cite{Atzmony1973,Koon1978,Kuzmin2001,Bowden2004,Martin2006},
which, although giving physical insight, requires parameters 
e.g.\ from experiment or point charge models,
which are difficult to fit.
For instance, Ref.~\citenum{Kuzmin2001} demonstrates how three
different sets of CF parameters can reproduce the
same experimental magnetostriction curve for DyFe$_2$.
First-principles calculations are free of these parameters,
but are often limited to describing stoichiometric
compounds at zero temperature.

Here, we combine non-empirical
first-principles calculations with the 
CF approach
in order to calculate the spin orientation diagram of Tb$_{1-x}$Dy$_x$Fe$_2$
and the critical concentration $x_c(T)$.
Our approach takes the recently-introduced yttrium-analogue method
of calculating CF coefficients within density-functional theory (DFT)~\cite{Patrick20192}---
which is numerically stable and avoids problems traditionally associated
with describing highly-correlated 4$f$ electrons in DFT---
and extends it to compute the phenomenological model parameters
associated with magnetostriction.
CF theory is then used to calculate the magnetocrystalline and magnetoelastic
energies associated with these localized RE-4$f$ electrons.
We further include the magnetostrictive contribution from
itinerant electrons  using the finite temperature DFT-based
formulation of the disordered local moment picture.
Our calculated spin orientation diagram reproduces experimental
measurements of the [111] and [100] easy directions over the full
range of temperatures and concentrations.
We find the critical concentration $x_c$
to be 0.78 at room temperature with magnetostrictions
$\lambda_{111}$=2700 and $\lambda_{100}$=-430~ppm,
close to the Terfenol-D values.

The rest of our manuscript is organized as follows.
Section~\ref{sec.methodology} describes the theory behind
our calculation of the spin orientation diagram.
In particular, we introduce the phenomenological expression
for the total energy as a function of magnetization direction
and strain, and discuss the magnetocrystalline and magnetoelastic
constants which enter this expression.
We review how the contribution to these constants from RE-4$f$ electrons
can be connected to crystal field coefficients,
and describe how these coefficients are obtained within DFT.
We also discuss the disordered local moment calculations used
to obtain the itinerant electron contribution and the temperature
dependence of the RE-4$f$ magnetic moments.
We then present our results in Sec.~\ref{sec.results}, consisting
of the calculated magnetocrystalline and magnetoelastic
constants of TbFe$_2$ and DyFe$_2$, and then the 
composition and temperature-dependent
spin orientation diagram, which is the main result of this work.
Finally in Sec.~\ref{sec.outlook} we outline future research directions.

\section{Methodology}
\label{sec.methodology}

\subsection{Spin orientation at zero temperature}

Our calculations are based on the 
following phenomenological expression for the energy of the crystal,
\begin{equation}
E(\bm{\hat{e}},\bm{\varepsilon}) = 
E_\mathrm{el}(\bm{\varepsilon})  +
E_\mathrm{RE}(\bm{\hat{e}},\bm{\varepsilon})
+
E_\mathrm{itin}(\bm{\hat{e}},\bm{\varepsilon})
\label{eq.Etot}
\end{equation}
which consists of a magnetization-independent elastic
energy $E_\mathrm{el}$, a contribution $E_\mathrm{RE}$
originating from the 4$f$ electrons localized   on 
the RE atoms, and $E_\mathrm{itin}$, which originates
from itinerant (delocalized) electrons.
$\bm{\varepsilon}$ represents the strain, 
with components written either in Cartesian form ($\varepsilon_{xx}, \varepsilon_{xy}$ etc.)
or as linear combinations of these ($\varepsilon^\alpha$, $\varepsilon^{\gamma i}$,
$\varepsilon^{\epsilon i}$), where $\alpha$, $\gamma$ and $\epsilon$ describe
homogeneous, tetragonal and shear strain modes, respectively~\cite{Clark1980}.
$\bm{\hat{e}}$ is a unit vector describing the orientation of the magnetization,
which can be alternatively expressed as
$\bm{\hat{e}}=(\cos\phi\sin\theta,\sin\phi\sin\theta,\cos\theta)$.
The equilibrium strain and magnetization state is taken to
be that which minimizes $E(\bm{\hat{e}},\bm{\varepsilon})$.

The magnetization of the entire crystal can be seen as the sum of 
individual contributions from local magnetic moments, where each
local moment with some magnitude $\mu$ is associated with a magnetic atom~\cite{Gyorffy1985}.
At zero temperature, the local moments form an ordered magnetic structure.
Raising the temperature introduces thermal disorder amongst
the local moments which generally weakens the overall magnetization,
until complete disorder is reached at the Curie temperature~\cite{Gyorffy1985}.
In the zero temperature case,
$\bm{\hat{e}}$ describes equivalently the orientation
of a particular local moment \emph{or} the orientation of the overall 
magnetization~\footnote{We note that in a ferrimagnet like REFe$_2$,
the itinerant electron magnetic sublattice may be oriented 
\emph{antiparallel} to $\bm{\hat{e}}$; however this detail does not affect our discussion,
since
$E_\mathrm{itin}(\bm{\hat{e}},\bm{\varepsilon}) = E_\mathrm{itin}(-\bm{\hat{e}},\bm{\varepsilon})$
 }.
However, this equivalence does not hold at finite temperature, where
the magnetic properties of the crystal are determined as an average
over the fluctuating local moments.
We concentrate initially on the zero temperature case.
The generalization to finite temperature is discussed in Sec.~\ref{sec.finiteT}.
We now discuss each term in equation~\ref{eq.Etot}:

\subsubsection{Elastic energy}
\label{sec.elastic}

The elastic energy  is quadratic in strain
and depends on the three elastic constants $c_{11}$, $c_{12}$ and 
$c_{44}$~\cite{Kittel1949,Clark1980}:
\begin{eqnarray}
E_\mathrm{el}(\bm{\varepsilon})&=&
\frac{c_{11}}{2} (\varepsilon_{xx}^2 + \varepsilon_{yy}^2 + \varepsilon_{zz}^2)
+ c_{12} (\varepsilon_{yy}\varepsilon_{zz} + \varepsilon_{zz}\varepsilon_{xx}
\nonumber \\
&& + \varepsilon_{xx}\varepsilon_{yy})+
\frac{c_{44}}{2}  (\varepsilon_{xy}^2 + \varepsilon_{yz}^2 + \varepsilon_{zx}^2)
\end{eqnarray}
Ideally we should calculate these constants from first principles.
However, even obtaining zero temperature elastic constants for the stoichiometric
end compounds TbFe$_2$ and DyFe$_2$ in DFT is not straightforward
due to the difficulty in treating the RE-4$f$ electrons~\cite{Bentouaf2016}.
Furthermore, the elastic constants are, in principle, dependent on composition
and temperature.
For simplicity we instead use a single set of elastic constant values of 
141, 65 and 49~GPa for $c_{11}$, $c_{12}$ and $c_{44}$, for all compositions
and temperatures.
These values were measured experimentally for Tb$_{0.3}$Dy$_{0.7}$Fe$_2$~\cite{Clark1980}.
We have tested the sensitivity of our results to this choice by calculating
spin orientation diagrams using different sets of elastic constant values
which were either obtained from  DFT or measured 
experimentally, for different compositions~\cite{Bentouaf2016,Moulay2013}.
The comparison is provided as an Appendix~\ref{sec.app}, and shows
the sensitivity to be very weak.

\subsubsection{RE-4$f$ electron energy}
The energy associated with the RE-4$f$ electrons 
$E_\mathrm{RE}(\bm{\hat{e}},\bm{\varepsilon})$
can be further partitioned
as
\begin{equation}
E_\mathrm{RE}(\bm{\hat{e}},\bm{\varepsilon}) = E_\mathrm{mca}(\bm{\hat{e}}) +
E_\mathrm{me}(\bm{\hat{e}},\bm{\varepsilon})
\label{eq.E4f}
\end{equation}
where the MCA energy $E_\mathrm{mca}(\bm{\hat{e}})$ depends only on the
orientation of the RE-4$f$ magnetic moment, and the magnetoelastic energy
$E_\mathrm{me}(\bm{\hat{e}},\bm{\varepsilon})$ couples this orientation to the strain.
The MCA energy can be written as
\begin{equation}
E_\mathrm{mca}(\bm{\hat{e}}) = \sum_{l=4,6} \mathcal{K}^{\alpha,l}S^{\alpha,l}(\bm{\hat{e}})
\label{eq.mca}
\end{equation}
where $\mathcal{K}^{\alpha,l}$ are the anisotropy constants
and $S^{X,l}$ are the symmetry basis functions, which are listed in Ref.~\citenum{Clark1980}
($X = \alpha, \gamma, \epsilon$).
$\mathcal{K}^{\alpha,l}$ are related to the more conventional anisotropy
constants $K_1$ and $K_2$ as $K_1 = -2(\mathcal{K}^{\alpha,4} + \frac{1}{22}\mathcal{K}^{\alpha,6})$
and $K_2 = \mathcal{K}^{\alpha,6}$.

The magnetoelastic energy $E_\mathrm{me}(\bm{\hat{e}},\bm{\varepsilon})$
is obtained as the direct product of strain and magnetization 
basis functions belonging to the same representation~\cite{Clark1980}:
\begin{eqnarray}
E_\mathrm{me}(\bm{\hat{e}},\bm{\varepsilon}) &=& 
\varepsilon^\alpha\sum_{l=4,6}\mathcal{B}^{\alpha,l}S^{\alpha,l}(\bm{\hat{e}})\nonumber \\
&&+\sum_{i=1,2} \varepsilon^{\gamma i}\sum_{l=2,4,6}\mathcal{B}^{\gamma,l}S^{\gamma,l}_i(\bm{\hat{e}}) \nonumber \\
&&
+\sum_{i=1,2,3} \varepsilon^{\epsilon i}\sum_{l=2,4,6,6'}\mathcal{B}^{\epsilon,l}S^{\epsilon,l}_i(\bm{\hat{e}})
\label{eq.me}
\end{eqnarray}
The coefficients
$\mathcal{B}^{X,l}$
are the magnetoelastic constants.
Note how the lower symmetry of the tetragonal or shear-strained structures
($\gamma$ or $\epsilon$) generates
new terms with an $l$=2 dependence on magnetization direction.

Evaluating $E_\mathrm{RE}(\bm{\hat{e}},\bm{\varepsilon})$ therefore
requires knowing the anisotropy and magnetoelastic constants
$\mathcal{K}^{\alpha,l}$ and 
$\mathcal{B}^{X,l}$.
We discuss the calculation of these constants within the
framework of the single-ion model and crystal field theory 
in Secs.~\ref{sec.singleion}, \ref{sec.CF} and \ref{sec.Yanalogue}.

\subsubsection{Itinerant electron energy}
\label{sec.itGd}

The remaining term $E_\mathrm{itin}(\bm{\hat{e}},\bm{\varepsilon})$ accounts
for the MCA and magnetoelastic contributions
to the energy not already included in the RE-4$f$ term,
i.e.\ those coming from itinerant electrons.
These itinerant electrons are mainly Fe-3$d$ in character,
with a lesser contribution from the RE-5$d$ 
electrons~\cite{Brooks19912}.
The relative importance of  $E_\mathrm{RE}$ and
$E_\mathrm{itin}$ to the magnetostriction
can be assessed by comparing TbFe$_2$ or DyFe$_2$
to their isostructural counterpart GdFe$_2$.
These three compounds have the same itinerant
electronic structure, and therefore should have comparable
$E_\mathrm{itin}$.
However, $E_\mathrm{RE}$ is zero in GdFe$_2$ 
due to the filled Gd-4$f$ spin subshell having zero
orbital moment~\cite{Kuzmin2008}.
Comparing the experimentally-measured
magnetostrictions of the different compounds,
we find TbFe$_2$ has a magnetostriction which is 50
times larger than GdFe$_2$~\cite{Clark1980},
showing that $E_\mathrm{RE}$ is the dominant contribution
to equation~\ref{eq.Etot}.
Nevertheless, for completeness we still include $E_\mathrm{itin}$
in our analysis.

In principle, $E_\mathrm{itin}(\bm{\hat{e}},\bm{\varepsilon})$
can be split into MCA and magnetoelastic contributions
as in equation~\ref{eq.E4f}, with a different set
of constants.
In practice (Sec.~\ref{sec.constants}), we find the MCA contribution
to be negligible, and also that it is sufficient
only to consider the $l=2$ term in the magnetoelastic
expansion.
We therefore have
\begin{equation}
E_\mathrm{itin}(\bm{\hat{e}},\bm{\varepsilon})
= \mathcal{B}^{\gamma,2}_{\mathrm{itin}} \sum_{i=1,2} \varepsilon^{\gamma i}S^{\gamma,2}_i(\bm{\hat{e}}) 
+\mathcal{B}^{\epsilon,2}_{\mathrm{itin}}\sum_{i=1,2,3} \varepsilon^{\epsilon i}S^{\epsilon,2}_i(\bm{\hat{e}})
\end{equation}
Due to their itinerant electron origin,
the constants
$\mathcal{B}^{\gamma,2}_{\mathrm{itin}}$ and
$\mathcal{B}^{\epsilon,2}_{\mathrm{itin}}$ cannot be obtained
from crystal field theory.
They are however amenable to treatment in the DFT-based
disordered local moment picture~\cite{Gyorffy1985,Marchant2019}.
We describe these calculations in Sec.~\ref{sec.itinerant}.

\subsection{Single-ion treatment of RE-4$f$ contribution and
modelling of alloys}
\label{sec.singleion}

We calculate the 
RE-4$f$ energy 
$E_\mathrm{RE}(\bm{\hat{e}},\bm{\varepsilon})$
within the single-ion model~\cite{Callen1966},
which has been used to great effect to understand
the behavior of RE-transition metal compounds
for many years~\cite{Kuzmin2008}.
In this model, the magnetic moment associated with the 4$f$ 
electrons localized on a particular RE ion behaves 
independently of its neighbours, which is a reasonable 
approximation~\cite{Kuzmin2001} given the highly-localized nature of these electrons
and the relatively weak RE-RE magnetic 
interactions measured in neutron scattering experiments~\cite{Loewenhaupt1996}.
The 4$f$ electrons localized at different RE sites experience the same
potential, which is an atomic-like central potential plus a contribution
from the surrounding crystal field.
The RE-4$f$ electrons also all experience an exchange field
originating from the itinerant electrons and possibly an external
magnetic field, which both drive magnetic order~\cite{Kuzmin2008}.

The crystal field is supposed to derive from the valence
electronic structure, and therefore is insensitive 
to (a) the orientations of surrounding RE-4$f$ localized moments,
and (b) the chemical species (Tb or Dy) of surrounding
RE ions (since these species have the same $6s^25d$ valence
electronic structure).
This latter aspect allows a simple treatment
of Tb-Dy alloying within the single-ion model, since each
RE ion is independent:
for a given composition Tb$_{1-x}$Dy$_x$Fe$_2$,
the RE-4$f$ energy per ion is a superposition
of the Tb and Dy contributions,
\begin{equation}
E_\mathrm{RE}(\bm{\hat{e}},\bm{\varepsilon}) = 
(1-x)E_\mathrm{Tb}(\bm{\hat{e}},\bm{\varepsilon}) + 
x E_\mathrm{Dy}(\bm{\hat{e}},\bm{\varepsilon}),
\label{eq.mix}
\end{equation}
where now $E_\mathrm{Tb}$ and $E_\mathrm{Dy}$ can
be seen as the RE-4$f$ energy contributions 
calculated for the end compounds TbFe$_2$ and DyFe$_2$,
respectively.
These end compounds each have their own set of
two anisotropy and nine magnetoelastic constants,
so to evaluate 
$E_\mathrm{RE}$ for an arbitrary $x$ we require
22 constants in total.

\subsection{RE anisotropy and magnetoelastic constants from crystal field theory}
\label{sec.CF}

In the single-ion central potential, the RE-4$f$ electrons
form atomic-like eigenstates
$|L,S,J,M_J\rangle$, where
$L$ and $S$ are determined by Hund's rules, 
$J = L + S$ for Tb and Dy, and 
$M_J = J, J-1,...,-J$~\cite{Griffithbook}.
Now, we should construct a Hamiltonian for the RE-4$f$ electrons
including the crystal, exchange and external fields,
and diagonalize it within the manifold of states with different $M_J$~\cite{Patrick20193}.
Without the crystal and external fields, the ground state will be
$|L,S,J,-J\rangle$, with the quantization axis (magnetic moment direction)
aligned with the exchange field.
Taking this axis as $\bm{\hat{z}}$, 
the RE-4$f$ electron density $\rho^{(\bm{\hat{z}})}_{4f}(\bm{r})$ 
associated with $|L,S,J,-J\rangle$
 is given by~\cite{Sievers1982}:
\begin{equation}
\rho^{(\bm{\hat{z}})}_{4f}(\bm{r}) = n^0_{4f}(r)  \sum_{l=2,4,6} \mathcal{A}_l \left(\frac{2l+1}{4\pi}\right)^{\frac{1}{2}} Y_{l0}(\bm{\hat{r}})
\label{eq.chargedens}
\end{equation}
Here, $n^0_{4f}(r)$ is the radial density calculated for the unperturbed central potential~\cite{Kuzmin2008},
and $Y_{lm}(\bm{\hat{r}})$ are complex spherical harmonics.
$\mathcal{A}_l$ are RE-dependent numerical
factors formed from $J$ and Stevens coefficients,
which for Tb$^{3+}$ 
are $\mathcal{A}_2$  = -(1/3),
$\mathcal{A}_4$  =  (1/11)
and 
$\mathcal{A}_6$  = -(5/429),
and for Dy$^{3+}$ are
$\mathcal{A}_2$  = -(1/3), 
$\mathcal{A}_4$  =  -(4/33), and
$\mathcal{A}_6$  =  (25/429)~\cite{Sievers1982,Stevens1952}.
The RE-4$f$ charge density $\rho^{(\bm{\hat{e}})}_{4f}(\bm{r})$ corresponding to a 
general magnetic moment direction $\bm{\hat{e}}$
is obtained from equation~\ref{eq.chargedens}
by making the substitution $ Y_{l0}(\bm{\hat{r}}) \rightarrow \sum_m  e^{-im\phi}d^{(l)}_{m0}(\theta)Y_{lm}(\bm{\hat{r}})$,
where the functions $d^{(l)}_{m0}(\theta)$ are equal to $[(l-m)!/(l+m)!]^{\frac{1}{2}} P^m_l(\cos\theta)$ and
$P^m_l(x)$ are the associated Legendre polynomials~\cite{Edmonds}.

\begin{figure}
\includegraphics[width=0.6\columnwidth]{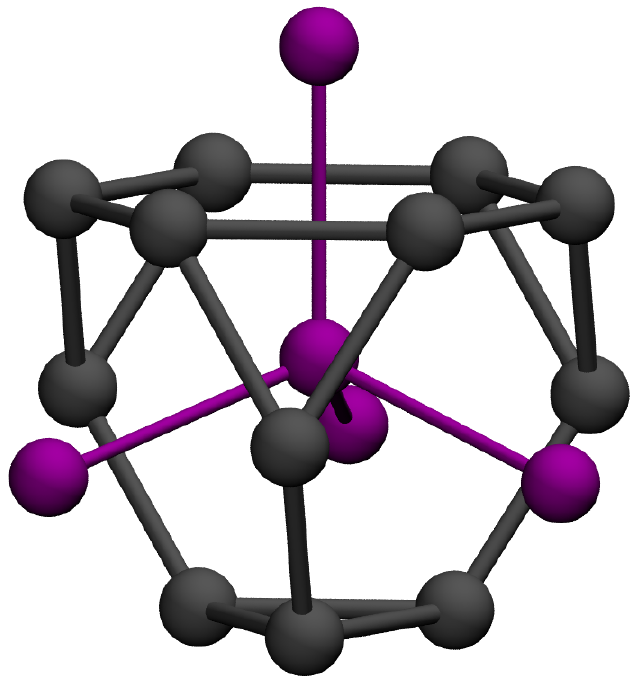}
\caption{
The local $T_d$ environment of the RE atom in the cubic Laves
phase, showing nearest neighbor (Fe, grey)  
and next-nearest neighbor (RE, purple) atoms.
The RE-RE bonds are oriented along the $\langle 111\rangle$ directions.
\label{fig.1}
}
\end{figure}

The crystal field (CF) characterizes the nonspherical components of the potential
at the RE site, $V(\bm{r}) = \sum_{lm} V_{lm}(r) Y_{lm}(\bm{\hat{r}})$.
If the exchange field is sufficiently strong compared to the CF, the latter
will not mix states of different $M_J$.
Then, the energy shift due to the CF is obtained from first order perturbation theory as
\begin{equation}
E_{4f}(\bm{\hat{e}}) = \sum_{l=2,4,6} \mathcal{A}_l \sum_m (-1)^m B_{l-m}e^{-im\phi}d^{(l)}_{m0}(\theta)
\label{eq.energy4f}
\end{equation}
where the CF coefficients~\cite{Patrick20192} have been introduced as:
\begin{equation}
B_{lm} = \left(\frac{2l+1}{4\pi}\right)^{\frac{1}{2}} \int r^2 n^0_{4f}(r)V_{lm}(r) dr.
\label{eq.Blm}
\end{equation}
For REFe$_2$ in the cubic Laves phase (Fig.~\ref{fig.1}), the RE atoms sit at sites with $T_d$ symmetry, so
the only nonzero CF coefficients which appear in equation~\ref{eq.energy4f} are $B_{40}$, $B_{4\pm4}$,
$B_{60}$ and $B_{6\pm4}$;
only $B_{40}$ and $B_{60}$ are independent~\cite{BradleyCracknell}.

Here we will assume that the exchange field is
strong enough that
$E_{4f}(\bm{\hat{e}})$ is given by equation~\ref{eq.energy4f},
and also that the exchange field and magnetization are isotropic.
Then, $E_{4f}$ is the only contribution to the energy which depends on the magnetization
angle.
At zero strain we can equate $E_{4f}$ and $E_{\mathrm{mca}}$ 
(equations~\ref{eq.mca} and $\ref{eq.energy4f}$) to obtain
\begin{equation}
\mathcal{K}^{\alpha,4} = \frac{5}{2} \mathcal{A}_4 B_{40}; \ \mathcal{K}^{\alpha,6} = \frac{231}{2} \mathcal{A}_6 B_{60}.
\label{eq.Ka46}
\end{equation}

Next, to obtain the magnetoelastic constants $\mathcal{B}^{X,l}$ we 
consider the modifications to the CF
coefficients when three different strain modes are applied:
 $\varepsilon_{xx} = \varepsilon_{yy} = \varepsilon_{zz} = 
\varepsilon_{I}$ (isotropic), $\varepsilon_{zz} = -2\varepsilon_{xx} = -2\varepsilon_{yy} = \varepsilon_T$
(tetragonal) and $\varepsilon_{xy} = \varepsilon_{yz} = \varepsilon_{zx} = 
\varepsilon_{S}$ (shear).
For the shear deformation it is convenient to work in a rotated co-ordinate system where
the $z$ axis coincides with the [111] direction.
Then, aside from altering the CF coefficients which are already nonzero in the
unstrained $T_d$ environment, the tetragonal and shear strains affect 
$E_{4f}(\bm{\hat{e}})$ in equation~\ref{eq.energy4f} by generating a nonzero $B_{20}$ coefficient.

Denoting the strain-induced shifts in CF coefficients as
$\Delta B_{lm}$, our calculations
(Sec.~\ref{sec.constants}) find that
these shifts can be described well by the linear relation
$\Delta B_{lm} =\frac{dB_{lm}}{d\varepsilon_X}\varepsilon_X$.
Inserting these relations into equation~\ref{eq.energy4f} and comparing to
equation~\ref{eq.me} gives each magnetoelastic constant
in terms of the strain derivative of a CF coefficient, for instance,
\begin{equation}
\mathcal{B}^{\gamma,2} = \frac{2}{3} \mathcal{A}_2 \frac{dB_{20}}{d\varepsilon_T}; \ \mathcal{B}^{\epsilon,2} = \mathcal{A}_2 \frac{dB_{20}}{d\varepsilon_S}.
\label{eq.Bg2Be2}
\end{equation}

\subsection{DFT calculation of CF coefficients}
\label{sec.Yanalogue}

Equations~\ref{eq.Ka46} and \ref{eq.Bg2Be2}
show how the anisotropy and magnetoelastic constants
can be obtained from the CF coefficients $B_{lm}$ and their
strain derivatives $dB_{lm}/d\varepsilon_X$.
These are the quantities which we calculate from first
principles within the yttrium-analogue method~\cite{Patrick20192}.
In this approach, the potential $V(\bm{r})$ which determines
the CF coefficients  is
calculated within DFT for the ``Y-analogue'' of
TbFe$_2$ or DyFe$_2$, which is YFe$_2$.
Specifically, the components $V_{lm}(r)$ in 
equation~\ref{eq.Blm} are found from the angular
decomposition of the self-consistent Kohn-Sham
potential calculated for the desired REFe$_2$
structure, where the RE is replaced with Y.

We have previously used the Y-analogue
method to calculate CF coefficients for various 
RE/transition-metal compounds~\cite{Patrick20192},
demonstrating its applicability to
describe temperature and pressure-induced
spin-reorientation transitions in
the RECo$_5$ compounds~\cite{Patrick20193,
Kumar2020,Kumar20202}.
Substituting Tb or Dy with Y
to calculate the crystal field is consistent with the assumptions
of the single-ion model~\cite{Callen1966}, 
namely that the CF depends on
the valence electronic structure 
and not on the RE-4$f$ electrons themselves.
Since the RE ions are in the 3$+$ state and therefore
are isovalent (two $s$ and a single $d$ electron),
we expect the CF of YFe$_2$ to be a good approximation for
TbFe$_2$ or DyFe$_2$.
Indeed, using the Y-analogue ensures that there is no double-counting
of the RE-4$f$ electrons in equation~\ref{eq.Blm}.
Any DFT implementation
can be used to calculate the CF coefficients, providing the valence
charge density is described accurately.

Equation~\ref{eq.Blm} also contains the RE-4$f$
electron density calculated for the unperturbed
central potential
$n^0_{4f}(r)$.
Previously we calculated $n^0_{4f}(r)$ within
self-interaction-corrected DFT~\cite{Lueders2005,Daene2009}
for a number of compounds and found that, for
a given RE element, it was highly insensitive
to the crystalline environment~\cite{Patrick20192}.
Therefore when calculating CF coefficients
we use the same previously calculated RE-dependent functions
($n^0_{4f,\mathrm{Tb}}(r)$ or $n^0_{4f,\mathrm{Dy}}(r)$
~\footnote{These functions were reported previously in Ref.~\cite{Patrick20192},
Fig.~1 and are available on request.})
for all strain states.

\subsection{Itinerant electron contribution}
\label{sec.itinerant}

The itinerant electrons are (by definition)
delocalized, and are responsible for generating the 
crystal field rather than simply being influenced by it.
Accordingly, the CF picture is not appropriate
to describe their contribution to the magnetostriction.
However, itinerant electron magnetism is amenable
to a fully first-principles treatment within DFT~\cite{Gyorffy1985}.
In Sec.~\ref{sec.itGd} we used GdFe$_2$ as a comparison system to 
understand the importance of $E_\mathrm{itin}(\bm{\hat{e}},\bm{\varepsilon})$
to the magnetostriction, since it has the same valence electronic structure
but zero CF contribution from the filled Gd-4$f$ spin subshell.
Building on this idea,  we take $E_\mathrm{itin}(\bm{\hat{e}},\bm{\varepsilon})$
to be the same for Tb$_{1-x}$Dy$_x$Fe$_2$ and GdFe$_2$, and calculate
the latter directly.
Similarly to using the Y-analogue, this approach avoids any double-counting
of the CF contribution.
However, using Gd rather than Y to calculate $E_\mathrm{itin}$
has the advantage of capturing any additional on-site polarization
of the valence electrons by the large spin moments possessed by
Gd, Tb and Dy~\cite{Patrick2018,Patrick20182}.

We calculate $E_\mathrm{itin}(\bm{\hat{e}},\bm{\varepsilon})$
for GdFe$_2$  using the same method demonstrated
recently for bcc Fe and Fe-Ga alloys~\cite{Marchant2019}.
This approach is a Green's function, multiple-scattering 
theory-based formulation of the 
disordered local moment picture within DFT (DFT-DLM~\cite{Gyorffy1985})
which as discussed in Sec.~\ref{sec.finiteT} allows
the treatment of finite temperature magnetic disorder.
The filled Gd-4$f$ spin subshell is treated efficiently 
using the local self-interaction correction (LSIC)~\cite{Lueders2005}.
Quantities related to the magnetic anisotropy are obtained
by solving the relativistic single-site scattering problem 
and applying the torque method~\cite{Staunton2006}. 
As described in Ref.~\cite{Marchant2019}, calculating
the derivative of the total energy with respect to magnetization
angle for different strain states allows the
anisotropy and magnetoelastic constants to be obtained.

\subsection{Generalization to finite temperature}
\label{sec.finiteT}

The methodology described above is sufficient to evaluate
equation~\ref{eq.Etot} assuming that all the individual
magnetic moments are ordered, corresponding to zero temperature.
At finite temperature $T$, equation~\ref{eq.Etot} takes a slightly
different form:
\begin{equation}
E(\bm{\hat{n}},\bm{\varepsilon},T) = 
E_\mathrm{el}(\bm{\varepsilon})  +
E_\mathrm{RE}(\bm{\hat{n}},\bm{\varepsilon},T)
+
E_\mathrm{itin}(\bm{\hat{n}},\bm{\varepsilon},T)
\label{eq.EtotT}
\end{equation}
The new quantity introduced is the unit vector $\bm{\hat{n}}$
which describes the orientation of the magnetization
of the entire crystal, and therefore represents an average over
the individual magnetic moments.
The degree of magnetic order is quantified through the temperature-dependent 
order parameters $m_{\mathrm{Tb}}$, $m_{\mathrm{Dy}}$ and $m_{\mathrm{itin}}$
which take values between 1 (zero temperature, fully ordered) and 0 (above
the Curie temperature, fully disordered).
The relationship between the orientation of the individual moments and their
average is given by, for instance, $
\langle \bm{\hat{e}_{\mathrm{Tb}}}\rangle_T =  m_{\mathrm{Tb}}(T)\bm{\hat{n}}$
where  $\langle \rangle_T$ denotes the
statistical mechanical thermal average taken (in this example)
over the individual moments of all Tb ions.
More generally, the finite and zero temperature energies in equations~\ref{eq.Etot}
and~\ref{eq.EtotT} are related simply as
$E(\bm{\hat{n}},\bm{\varepsilon},T) = \langle E(\bm{\hat{e}},\bm{\varepsilon}) \rangle_T$.

Evaluating the thermal average
$\langle \rangle_T$ requires a model for the statistical mechanics
of the magnetic moments.
The DFT-DLM framework employs a Heisenberg-like Hamiltonian 
for this purpose~\cite{Gyorffy1985}.
The probability that a moment is aligned along a direction $\bm{\hat{e}}$ 
at $T$ is given by $P_{\bm{\hat{n}}}(\bm{\hat{e}}) \propto \exp[\beta h\bm{\hat{n}}\cdot\bm{\hat{e}}]$,
where $1/\beta = k_BT$.
The Weiss field felt by each local moment $h(T)$ is determined self-consistently
from DFT-DLM calculations at a given temperature using the iterative scheme
described in Ref.~\cite{Patrick2017}.
The self-consistency condition ensures (a) that the free
energy is minimized, and (b) that the model approximates the true
statistical mechanics of the moments as closely as possible~\cite{Gyorffy1985}.
Each crystallographically inequivalent magnetic atom (Tb, Dy and Fe) 
experiences its own Weiss field, and within the model
the order parameter and Weiss fields are linked according to
(again taking Tb as an example):
\begin{equation}
m_{\mathrm{Tb}}(T) = \coth(\beta h_{\mathrm{Tb}}(T)) - \frac{1}{\beta h_{\mathrm{Tb}}(T)}
\end{equation}

\subsubsection{Thermally-averaged rare earth contribution}

Recalling that in the crystal field picture the CF is
independent of the RE moment orientations, and that the anisotropy
and magnetoelastic constants are determined by the CF coefficients, 
the thermal average of the RE contribution is determined by
solely by the average of
the symmetry basis functions, e.g.
\begin{equation}
E_\mathrm{mca}(\bm{\hat{n}},T) = \sum_{l=4,6} \mathcal{K}^{\alpha,l}\langle S^{\alpha,l}(\bm{\hat{e}})\rangle_T.
\label{eq.mcaT}
\end{equation}
Due to the local nature of the probability function $P_{\bm{\hat{n}}}(\bm{\hat{e}})$,
the general arguments of Callen and Callen~\cite{Callen1966} can be used to show
\begin{equation}
\langle S^{X,l}(\bm{\hat{e}}) \rangle_T = f_l(m) S^{X,l}(\bm{\hat{n}})
\end{equation}
where the functions $f_l(m)$ depend on $m$ as $m^{\frac{l(l+1)}{2}}$
and $m^l$ at low and high temperature respectively~\cite{Callen1966}.
Then, the explicit expression for the RE contribution at finite temperature is
\begin{eqnarray}
E_\mathrm{RE}(\bm{\hat{n}},\bm{\varepsilon},T) &=&
\sum_{l=4,6}  \mathcal{K}^{\alpha,l}_{\mathrm{RE}}(T)S^{\alpha,l}(\bm{\hat{n}}) \nonumber \\
&&
+\varepsilon^\alpha\sum_{l=4,6}\mathcal{B}^{\alpha,l}_{\mathrm{RE}}(T)S^{\alpha,l}(\bm{\hat{n}})\nonumber \\
&&+\sum_{i=1,2} \varepsilon^{\gamma i}\sum_{l=2,4,6}\mathcal{B}^{\gamma,l}_{\mathrm{RE}}(T)S^{\gamma,l}_i(\bm{\hat{n}}) \nonumber \\
&&
+\sum_{i=1,2,3} \varepsilon^{\epsilon i}\sum_{l=2,4,6,6'}\mathcal{B}^{\epsilon,l}_{\mathrm{RE}}(T)S^{\epsilon,l}_i(\bm{\hat{n}})
\nonumber \\
\label{eq.ERET}
\end{eqnarray}
where the finite and zero-temperature constants are simply related by  $f_l$,
\begin{eqnarray}
\mathcal{K}^{\alpha,l}_{\mathrm{RE}}(T) = \mathcal{K}^{\alpha,l}_{\mathrm{RE}} f_l(m_{\mathrm{RE}}(T)) \nonumber \\
\mathcal{B}^{X,l}_{\mathrm{RE}}(T) = \mathcal{B}^{X,l}_{\mathrm{RE}} f_l(m_{\mathrm{RE}}(T)) 
\end{eqnarray}
and the RE subscript has been inserted as a reminder that the constants and
order parameters are calculated either for TbFe$_2$ or DyFe$_2$.
The RE contribution for the Tb$_{1-x}$Dy$_x$Fe$_2$ alloy is obtained through
the same linear mixing as at zero temperature, as in equation~\ref{eq.mix}.

The temperature dependence of 
$E_\mathrm{RE}(\bm{\hat{n}},\bm{\varepsilon},T)$ is therefore fixed
by the order parameter dependences $m_{\mathrm{Tb}}(T)$
and $m_{\mathrm{Dy}}(T)$, which we
determine through finite-temperature, LSIC
DFT-DLM calculations on TbFe$_2$ and DyFe$_2$.
The calculations were performed according to the methodology described
in detail in Ref.~\cite{Patrick20182}, and the reader is referred there
for a more complete discussion of the underlying theory and technical
details of the DFT-DLM scheme.

\subsubsection{Thermally-averaged itinerant electron contribution}

Performing the thermal average on 
$E_\mathrm{itin}$ gives, in analogy with equation~\ref{eq.ERET},
\begin{eqnarray}
E_\mathrm{itin}(\bm{\hat{n}},\bm{\varepsilon},T)
&=&\mathcal{B}^{\gamma,2}_{\mathrm{itin}}(T) \sum_{i=1,2} \varepsilon^{\gamma i}S^{\gamma,2}_i(\bm{\hat{n}}) \nonumber \\
&&+\mathcal{B}^{\epsilon,2}_{\mathrm{itin}}(T)\sum_{i=1,2,3} \varepsilon^{\epsilon i}S^{\epsilon,2}_i(\bm{\hat{n}})
\end{eqnarray}
The finite temperature magnetoelastic constants are obtained from DFT-DLM
calculations on GdFe$_2$,
which
give directly the temperature dependence of  $\mathcal{B}^{X,2}_{\mathrm{itin}}$.
As found previously for bcc Fe~\cite{Marchant2019}, 
the $\mathcal{B}^{X,2}_{\mathrm{itin}}$ constants 
do not follow an $f_2(m_{\mathrm{itin}})$ dependence on
the order parameter.
This observation reflects the itinerant origin
of the magnetic anisotropy, compared to the single-ion
description of the RE moments~\cite{Staunton2006}.

\subsubsection{The need for a phenomenological model}

It is reasonable to ask, given that LSIC DFT-DLM calculations
can be used to obtain the itinerant electron magnetostriction
and also the temperature dependence of the RE order parameters
in TbFe$_2$ and DyFe$_2$, why we should not perform
the entire calculation in the DFT-DLM framework without any
reference to crystal field theory.
The technical difficulty is that the DFT-DLM calculations
are performed within the atomic sphere approximation (ASA)~\cite{Gyorffy1979},
which means that nonspherical components of the potential
at the RE site (i.e.\ the crystal field) are poorly described
in the DFT-DLM calculation of the RE anisotropy.
As a result, a separate treatment of the CF is required.
In turn, it is important that
the calculated energy contribution associated with the
itinerant electron anisotropy is free of any contribution
from the localized RE-4$f$ electrons interacting with the CF,
otherwise this contribution would be counted both in $E_\mathrm{itin}$
and $E_\mathrm{RE}$ in equation~\ref{eq.Etot}.
Calculating the itinerant contribution for GdFe$_2$, 
which has no CF anisotropy, ensures that this is the case.
Similarly, the assumptions of the CF model mean that the CF coefficients
themselves should not depend on the asphericity of the RE-4$f$ electrons.
This requirement is satisfied by 
using the Y-analogue model, where the RE-4$f$ electrons do not
enter the calculation  of the CF potential at all~\cite{Patrick20192}.
These same considerations led us to adopt a similar scheme
in the calculation of finite temperature anisotropy of the RCo$_5$
compounds~\cite{Patrick20193}.

\subsection{Computational details}
Crystal field coefficients were calculated for YFe$_2$ 
within the projector-augmented formulation of DFT as implemented
in the GPAW code~\cite{Enkovaara2010},
using the local spin-density approximation (LSDA) for exchange
and correlation~\cite{Vosko1980}.
A plane wave basis set with a 1200~eV energy cutoff and a 20$\times$20$\times$20 $k$-point
sampling was used, as in Ref.~\citenum{Patrick20192}.
A lattice constant of 7.341~\AA \ was used throughout for the equilibrium (cubic) structure, 
which is the experimentally-measured value for TbFe$_2$ at room temperature~\cite{AndreevHMM};
the value for DyFe$_2$ is very similar (7.338~\AA).
The dependence of the order parameters on temperature were calculated
within DFT-DLM~\cite{Gyorffy1985} with the LSIC applied~\cite{Patrick20182},
using the ASA with Wigner-Seitz radii of 1.90~\AA \ 
for the RE atoms, with angular momentum expansions truncated at $l_\mathrm{max} = 3$.
The same computational setup was used to calculate the temperature-dependent 
magnetoelastic constants associated with the itinerant electrons for GdFe$_2$,
using the torque method as described in Refs.~\citenum{Marchant2019} and \citenum{Staunton2006}.

\section{Results}
\label{sec.results}

\subsection{Anisotropy and magnetoelastic constants}
\label{sec.constants}
\begin{table*}
\centering
\begin{tabular}{lccccccccccccc}
\hline
&
$\mathcal{K}^{\alpha,4}$&
$\mathcal{K}^{\alpha,6}$&
$\mathcal{B}^{\alpha,4}$&
$\mathcal{B}^{\alpha,6}$&
$\mathcal{B}^{\gamma,2}$&
$\mathcal{B}^{\gamma,4}$&
$\mathcal{B}^{\gamma,6}$&
$\mathcal{B}^{\epsilon,2}$&
$\mathcal{B}^{\epsilon,4}$&
$\mathcal{B}^{\epsilon,6}$&
$\mathcal{B}^{\epsilon,6'}$&
$\mathcal{B}^{\gamma,2}_\mathrm{itin}$& 
$\mathcal{B}^{\epsilon,2}_\mathrm{itin}$\\
\hline
TbFe$_2$ & 14.10 & 11.20 & -22.87 & -27.88 & 74.69 & 28.08 &  6.96 & -844.24 & 258.17 &  7.22 & -4.20 & -7.25 & 33.24 \\
DyFe$_2$ & -17.34 & -47.52 & 28.14 & 116.89 & 77.12 & -30.87 & -29.43 & -794.88 & -307.96 & -33.07 & 17.79 & -7.25 & 33.24
\end{tabular}
\caption{Anisotropy and magnetoelastic constants in MJm$^{-3}$ calculated for
TbFe$_2$ and DyFe$_2$.
\label{tab.coffs}
}
\end{table*}

We previously reported Y-analogue 
calculations of the CF coefficients of TbFe$_2$ and DyFe$_2$~\cite{Patrick20192}.
The values of $\mathcal{K}^{\alpha,4}$ and  $\mathcal{K}^{\alpha,6}$ calculated from 
equation~\ref{eq.Ka46} are given in Table~\ref{tab.coffs}.
Importantly, due to the differences in $\mathcal{A}_4$ and  $\mathcal{A}_6$ for Tb$^{3+}$ and Dy$^{3+}$,
 $\mathcal{K}^{\alpha,l}$ have opposite signs for TbFe$_2$ and DyFe$_2$ so favor
different magnetization directions.
From the linear mixing of equation~\ref{eq.mix}, we note that a Dy content
of $x=0.45$ would lead to a zero value of $\mathcal{K}^{\alpha,4}$.

\begin{figure}
\includegraphics[width=\columnwidth]{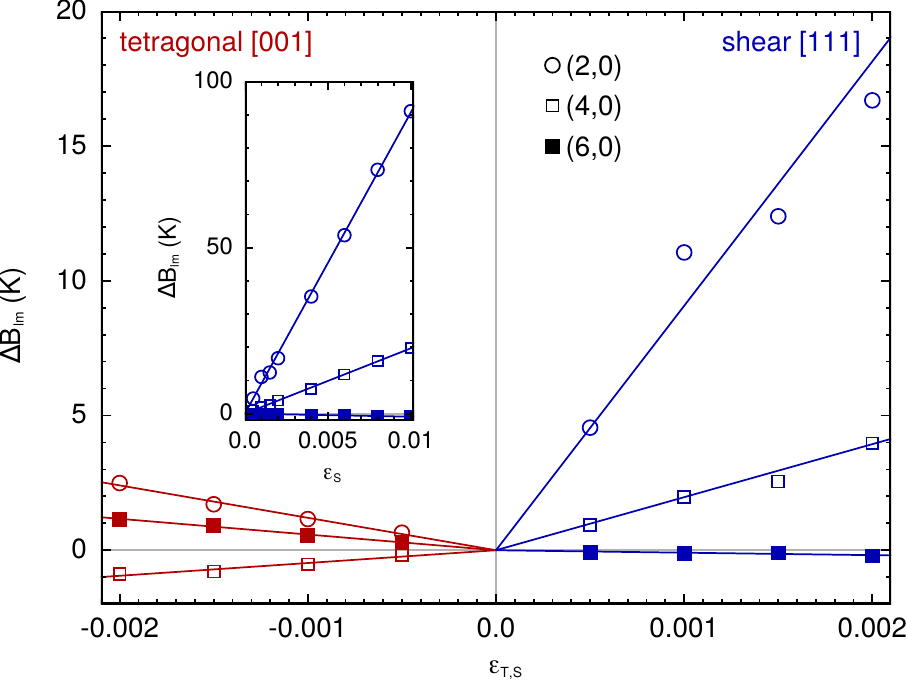}
\caption{Change in crystal field coefficients $\Delta B_{lm}$ for TbFe$_2$ for different $(l,m)$ 
with a shear strain $\varepsilon_S$ (blue) or a tetragonal strain $\varepsilon_T$ (red) applied.
The inset shows $\Delta B_{lm}$ for a larger variation in $\varepsilon_S$.
The straight lines are fits to the calculations.
\label{fig.2}
}
\end{figure}

Now considering the magnetoelastic constants associated with the RE,
in Fig.~\ref{fig.2} we plot the strain-induced change in the CF coefficients
$\Delta B_{lm}$ for TbFe$_2$, for $(l,m)$ = (2,0), (4,0) and (6,0).
We show  $\Delta B_{lm}$ for both tetragonal ($\varepsilon_T$) and shear ($\varepsilon_S$) 
strains.
Following convention, we divide the CF coefficients by $k_B$ so that the
quantities have dimensions of temperature.

Although there is some numerical noise in $\Delta B_{20}$ evident for small shear strains $\varepsilon_S$, 
$\Delta B_{lm}$ is linear in $\varepsilon$ over the range of strains considered.
Indeed, extending the calculations to larger shear strains confirms this linear relation
out to at least $\varepsilon_S = 0.01$ (inset of Fig.~\ref{fig.2}).
Then, the most striking feature of Fig.~\ref{fig.2} is the strong dependence of $B_{20}$
on $\varepsilon_S$.
At $\varepsilon_S$ = 0.002, $\Delta B_{20}$ is 17~K, compared to 2~K 
for $\varepsilon_T$ at -0.002.
The corresponding difference between shear and tetragonal strains
is much reduced at larger $(l,m)$ values, with
$\Delta B_{40}$ = 4~K and -1~K and $\Delta B_{60}$ = 0~K and 1~K 
respectively.

Converting these derivatives into magnetoelastic constants
through relations like equation~\ref{eq.Bg2Be2} gives the values
shown in Table~\ref{tab.coffs}.
The large value of $\frac{dB_{20}}{d\varepsilon_S}$ is reflected
in the coefficient
$\mathcal{B}^{\epsilon,2}$, which is an order of magnitude larger than
$\mathcal{B}^{\gamma,2}$.
Since  $\mathcal{B}^{\epsilon,2}$ is negative, this term will favor
positive strains along [111].
Furthermore, $\mathcal{B}^{\epsilon,2}$ is the same sign for both TbFe$_2$
and DyFe$_2$, since 
$\mathcal{A}_2$ is identical for Tb$^{3+}$ and Dy$^{3+}$~\cite{Sievers1982,Stevens1952}.
Therefore,
unlike $\mathcal{K}^{\alpha,4}$,
 there is no  cancellation of 
$\mathcal{B}^{\epsilon,2}$ in the alloy.
It is this aspect which allows 
Tb$_{1-x}$Dy$_x$Fe$_2$ to have simultaneously a large
magnetostriction and small anisotropy.

Now considering the itinerant electrons, 
our DFT-DLM calculations on GdFe$_2$ find
the contribution to the MCA to be negligible (of order 1~Jm$^{-3}$).
The magnetoelastic constants are more significant, and their zero temperature
values are given in Table~\ref{tab.coffs} (we stress again that their 
temperature dependence is more complicated than $f_l(m)$)~\cite{Marchant2019}.
The magnetoelastic contribution is well described by constants with $l=2$ only.
$\mathcal{B}^{\gamma,2}_\mathrm{Fe}$  and $\mathcal{B}^{\epsilon,2}_\mathrm{Fe}$
are calculated to have the same sign as observed experimentally for bcc Fe~\cite{Wedler2000},
but their magnitudes are enhanced (-7.1 and
33~MJm$^{-3}$).
However, the itinerant electrons still contribute 
much less than the RE at all of the temperatures considered here.

\subsection{Easy directions and magnetostrictions at zero temperature}

Using the constants reported in Table~\ref{tab.coffs}
we can construct the phenomenological energy for an arbitrary strain, magnetization
and composition.
Considering the zero temperature case first (equation~\ref{eq.Etot}),
minimizing $E(\bm{\hat{e}},\bm{\varepsilon})$ with respect to magnetization
direction and strain for the end compounds TbFe$_2$ and DyFe$_2$ finds
easy directions of [111] and [100]
respectively.
The calculated fractional changes in length along [111] and
[100] for TbFe$_2$ and DyFe$_2$ are 
$\lambda_{111}^{\mathrm{TbFe_2}}$ = 5200~ppm and
$\lambda_{100}^{\mathrm{DyFe_2}}$ = -780~ppm at 0~K.
Comparing to experimentally-measured values
of 4400 and -70~ppm~\cite{Clark1977} shows
correct qualitative behaviour and numerical agreement
within $\sim$1000~ppm, or 0.1\% strain; 
in relative terms, the agreement for
$\lambda_{100}^{\mathrm{DyFe_2}}$ is less good
than for TbFe$_2$.

Now considering the alloy through equation~\ref{eq.mix}
we find a [111] easy direction for all values of $x$ 
below $x_c = 0.56$, above which the easy direction
switches abruptly to [100].
This is some way off the experimental optimal 
concentration of $x = 0.73$, but we have not yet included
temperature effects.
It is also interesting to recompute the magnetization direction
ignoring the magnetoelastic contribution to the energy.
Then, $x_c$ is found to be 0.45, the same value which
cancelled  $\mathcal{K}^{\alpha,4}$.

\subsection{Spin orientation diagram}
\begin{figure}
\includegraphics[width=\columnwidth]{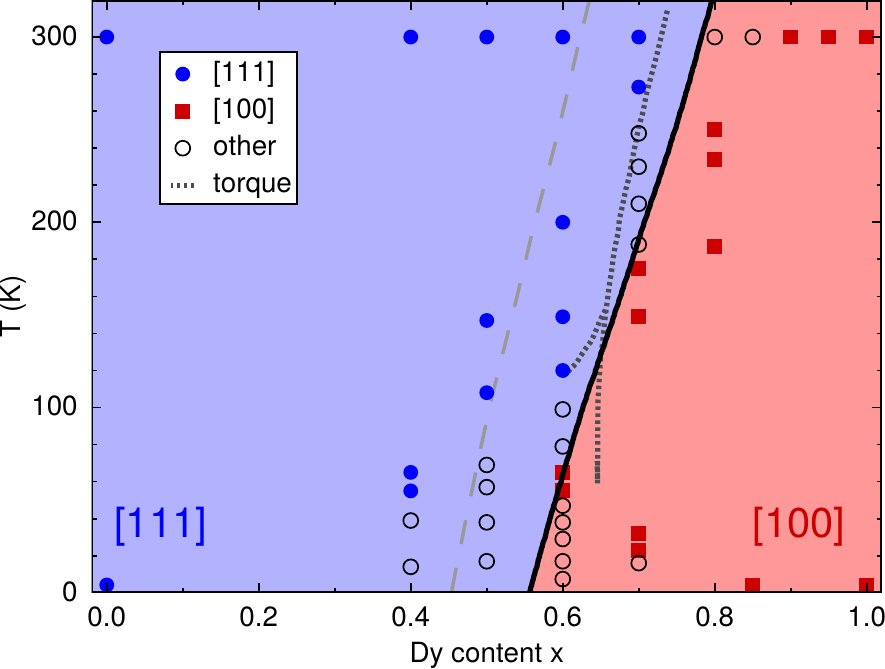}
\caption{
The easy direction of magnetization of Tb$_{1-x}$Dy$_x$Fe$_2$,
calculated by minimizing 
$E(\bm{\hat{n}},\bm{\varepsilon},x,T)$
(red and blue shaded regions).
The symbols are experimental measurements of the easy direction using M\"ossbauer
spectroscopy~\cite{Atzmony1973,Atzmony1977}.
The dotted lines mark the boundaries between different magnetization directions
extracted from torque magnetometry~\cite{Williams1980}, where above 150~K
the boundary is between [111] and [100], and below encloses a region of intermediate
magnetization direction.
The dashed line is the [111]/[100] boundary obtained 
by minimizing $E_\mathrm{MCA}$ only.
\label{fig.3}
}
\end{figure}
We now consider finite temperature, and minimize
$E(\bm{\hat{n}},\bm{\varepsilon},T)$ (equation~\ref{eq.EtotT})
for a grid of $(x,T)$ values.
The resulting spin orientation diagram is shown in Fig.~\ref{fig.3}.
As at zero temperature, the easy directions are found either 
to be [111] or [100] (blue or red regions), and
increased Dy content favours [100] magnetization.
However, at higher temperatures more Dy is required to maintain the [100]
magnetization, i.e.\ $x_c$ increases with temperature.

The reason for the increase in $x_c$ is due to the behavior of the RE 
order parameters with temperature.
Our DFT-DLM calculations  find that the
Dy order parameter $m_{\mathrm{Dy}}$ decreases
more quickly with $T$ than $m_{\mathrm{Tb}}$,
something which can also be inferred from experimental
magnetization measurements~\cite{Clark1980}.
This behavior can be understood
as the lower spin moment of Dy weakening the exchange
interaction~\cite{Brooks19912}.
Since $\mathcal{K}^{\alpha,4}$ is highly sensitive to $m$ ($\sim m^{10}$, thanks
to $f_4(m)$),
more Dy is required at higher temperatures to maintain the [100] magnetization.

Our calculated value of $x_c$ at 300~K is  $x_c=0.78$.
At this concentration we calculate magnetostrictions
of $\lambda_{111}$=2700 and $\lambda_{100}$=-430~ppm.
As at zero temperature with the end compounds, the
calculated values are
within $\sim$1000~ppm of the experimental ones, as measured
at 300~K for Terfenol-D~\cite{Abbundi1977}.

Like for the zero temperature case, we also calculated the spin orientation
ignoring the magnetoelastic terms in the energy.
The boundary between the [111] and [100] easy directions in this case
is shown as the grey dashed line in Fig.~\ref{fig.3}.
The shifted line can be understood from Fig.~\ref{fig.2} and surrounding
discussion: $\mathcal{B}^{\epsilon,2}$ is large,
so while the magnetization points along [111] the material can save energy 
by distorting.
Switching off the magnetoelastic contribution 
reduces the region where [111] magnetization is favorable, so less
Dy is required to make the transition to [100].

Figure~\ref{fig.3} also shows experimental measurements of the easy magnetization
direction obtained from M\"ossbauer spectroscopy~\cite{Atzmony1973,Atzmony1977},
and torque magnetometry measurements of the $(x,T)$ boundaries between different 
magnetization orientations~\cite{Williams1980}.
Our calculations agree with all of the measurements of the [111] and [100]
easy directions across different temperatures and compositions (no red symbols
appear on blue, and vice versa).
However, the open circles in Fig.~\ref{fig.3}
are measurements where the magnetization points along [$u$$v$0] or [$u$$v$$w$]
rather than [111] or [100]~\cite{Atzmony1977}.
Our calculations do not capture these intermediate directions, as we shall
discuss in the concluding section.

\section{Outlook}
\label{sec.outlook}

We first return to the original question of our work concerning 
Terfenol-D's optimum dysprosium content, $x$=0.73.
Our calculations actually find that the entire composition range of
Tb$_{1-x}$Dy$_x$Fe$_2$ is remarkable for having highly 
anisotropic magnetostrictions.
For instance, we find that the end 
compounds have $\lambda_{111}^{\mathrm{DyFe_2}}$ = 5640~and
$\lambda_{100}^{\mathrm{TbFe_2}}$ = -970~ppm at 0~K
(compare to
$\lambda_{111}^{\mathrm{TbFe_2}}$ = 5200~ppm and
$\lambda_{100}^{\mathrm{DyFe_2}}$ = -780~ppm reported above).
However, what is critical for applications is the ability to
rotate the magnetization direction at small fields~\cite{Clark1980},
i.e.\ a small MCA, which is achieved at $x_c$ where the easy
direction switches.
Our calculated value of $x_c$=0.78 at 300~K rationalizes
the experimentally-determined critical concentration from
first principles.
We stress that we get a very different value if we ignore 
temperature ($x_c$=0.56) or magnetostriction ($x_c$=0.62).

Interestingly our calculations have not captured a more subtle
feature of the spin orientation diagram, which is the presence of
[$u$$v$0] or [$u$$v$$w$] easy magnetization directions
(open circles in Fig.~\ref{fig.3})~\cite{Atzmony1977}.
The reason for this discrepancy is in
our first-order treatment of the CF, which
generates terms up to $l=6$ in equation~\ref{eq.Etot}.
In order to describe [$u$$v$0] or [$u$$v$$w$] easy directions, 
the energy must contain terms with larger $l$~\cite{Martin2006,Atzmony1976}.
To proceed, we should go beyond the
first-order perturbative treatment of the CF (equation~\ref{eq.energy4f})
and instead construct the full RE-4$f$ Hamiltonian including the CF potential
and the exchange field, and diagonalize it within the $M_J$ manifold~\cite{Patrick20193}.
A complete treatment would map out the strain dependence of all terms
within the Hamiltonian.
This approach could potentially find intermediate easy directions
and also allow us to calculate the dependence
of Tb$_{1-x}$Dy$_x$Fe$_2$ magnetostriction on the external field.
Our test calculations using a finite exchange field have indeed found
intermediate easy directions for small $T$ and  $x \sim$ 0.5,
indicating that this is a promising direction for future work.

A further refinement is to account for internal distortions within
the unit cell.
Indeed, the classic work of Cullen and Clark~\cite{Cullen1977} argued that the internal
distortion could provide the key to explaining the huge anisotropy in magnetostriction
between the [111] and [100] directions.
However, as was shown by the zero temperature calculations of Ref.~\cite{Buck1999} 
and reiterated here, $\lambda_{111}$ is
found to be much larger than $\lambda_{100}$ even when no internal distortions are
taken into account.
Our test calculations of the CF coefficients along different
frozen phonon modes have found the variation to be small compared to applying a global strain.
However, the (zero temperature) calculations of Ref.~\cite{Buck1999} did find a reduction
in $\lambda_{111}^{\mathrm{TbFe_2}}$ of 1300~ppm when they included an internal distortion, which would bring
our value closer to experiment.
Therefore, it is important to investigate the inclusion of all possible distortions and couplings at a consistent level.

An additional question concerns the use of the single-ion approximation
(e.g.\ equation~\ref{eq.mix}).
This approximation is generally understood to work very
well for rare-earth/transition-metal magnets like REFe$_2$~\cite{Kuzmin2008}.
However, it is reasonable to ask to what extent the crystal field parameters and the exchange
field at the RE site might be influenced by fluctuations in its surroundings,
including those caused by other RE atoms.
Employing our methodology on supercells incorporating such fluctuations will allow
this question to be addressed.

Going beyond Terfenol-D,
having validated the methodology we can now evaluate other
materials' magnetostrictive properties, ideally with reduced RE content.
The ability to calculate phase boundaries
is of particular interest to the design of multiferroic architectures,
where working at such boundaries will maximise the response~\cite{Li20182}.
For instance, we could easily simulate epitaxial strain by adding additional strain
to our calculations or, more ambitiously, model the explicit effects of the interface on the CF.
Intriguingly, the calculations also show that there exists a basic property of the Laves phase
structure, perhaps the orientation of RE-RE bonds, which makes the CF highly sensitive
to shear strain.
Elucidating this mechanism could help design more magnetostrictive materials.

\begin{acknowledgments}
The present work forms part of the PRETAMAG project,
funded by the UK Engineering and Physical Sciences Research
Council, Grant No. EP/M028941/1.
\end{acknowledgments}

\appendix*
\section{Elastic constants}
\label{sec.app}

\begin{table}
\begin{tabular}{rccc}
\hline
&$c_{11}$&$c_{12}$&$c_{44}$\\
\hline
Tb$_{0.3}$Dy$_{0.7}$Fe$_2$, exp.~\cite{Clark1980}& 141& 65& 49 \\
DyFe$_2$, exp.~\cite{Clark1980}& 146& 68& 47 \\
TbFe$_2$, calc.~\cite{Bentouaf2016}& 197& 112& 84 \\
YFe$_2$, calc.~\cite{Moulay2013}& 206& 132& 50 \\
\end{tabular}
\caption{Elastic constants, in GPa either measured
experimentally (exp.)
or calculated (calc.) for different compounds.
\label{tab.elastic}
}
\end{table}

In our calculations of the elastic energy (Sec.~\ref{sec.elastic})
we used the values of the elastic constants
$c_{11}$, $c_{12}$ and $c_{44}$
measured experimentally~\cite{Clark1980} 
for Tb$_{0.3}$Dy$_{0.7}$Fe$_2$
for all compositions and temperatures.
Here we illustrate the effect on the spin orientation diagram 
of using different values for these constants.
Table~\ref{tab.elastic} lists elastic constants
either measured experimentally for
Tb$_{0.3}$Dy$_{0.7}$Fe$_2$ and DyFe$_2$~\cite{Clark1980},
or calculated within DFT for TbFe$_2$ and YFe$_2$~\cite{Bentouaf2016,Moulay2013}.
For the DFT calculations a generalized-gradient approximation (GGA)
was used for the exchange correlation.
We include YFe$_2$ due to it having the same valence
electronic structure.

\begin{figure}
\includegraphics[width=\columnwidth]{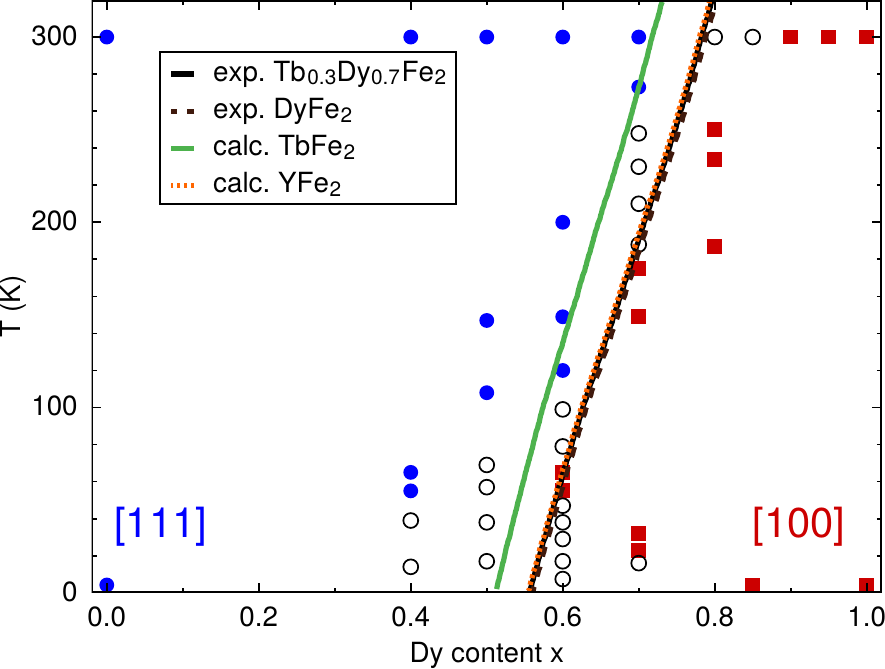}
\caption{The spin orientation diagram of 
Tb$_{1-x}$Dy$_x$Fe$_2$ calculated with different
sets of elastic constants.
The same experimental data is shown as in Fig.~\ref{fig.3}.
The diagonal lines represent the boundaries between
[111] and [100] directions of magnetization for
the different sets of elastic constants listed
in Table~\ref{tab.elastic}.
\label{fig.app1}
}
\end{figure}

We recalculated the spin orientation diagram for each
set of constants and show the result in Fig.~\ref{fig.app1}.
The qualitative structure of the diagram for each set
of constants is identical, consisting of a single
boundary between [111] and [100] easy directions.
Quantitatively, the three sets of elastic constants
corresponding to 
Tb$_{0.3}$Dy$_{0.7}$Fe$_2$ and DyFe$_2$ (experimental)
and YFe$_2$~\cite{Bentouaf2016,Moulay2013} (calculated)
give effectively identical boundaries.
Using the elastic constants calculated for TbFe$_2$
shifts the critical concentration $x_c$ down by approximately
0.05, such that $x_c = 0.51 $ at 0~K and $x_c = 0.72$ at
300~K.
Examining Table~\ref{tab.elastic} would indicate that the
critical concentration is most sensitive to $c_{44}$, which
is reasonable given crucial role played by the large [111] 
magnetostriction.

We note that using the elastic constants calculated for TbFe$_2$ brings
the room temperature critical concentration to within 0.01
of the experimental Terfenol-D value.
However, since it is not clear that a GGA treatment is sufficiently
accurate to describe the Tb-4$f$ electrons~\cite{Bentouaf2016,Patrick20182},
in this work we prefer to use experimental values
for the elastic constants.
Furthermore, the effectively identical results
for Tb$_{0.3}$Dy$_{0.7}$Fe$_2$ and DyFe$_2$, and the weak
sensitivity to $c_{ij}$ in general, justifies
the use of a single set of elastic constants for
the entire spin orientation diagram.

%\bibliography{papers}


\end{document}